\documentclass[aps,prd,twoside,twocolumn,superscriptaddress,floatfix,nofootinbib,showpacs]{revtex4-1}
\usepackage{amsmath}

\usepackage{bm}
\usepackage{xcolor}
\usepackage{hyperref}
\usepackage{url}
\usepackage{graphicx}
\usepackage[large]{subfigure}
\usepackage{amssymb}
\usepackage[amssymb]{SIunits}

\usepackage{natbib}
\usepackage{multirow}

\newcommand{\aap}{Astron. Astrophys.}

\newcommand{\jcap}{J. Cosmol. Astropart. Phys.}

\newcommand{\physrep}{Phys. Rep.}

\newcommand{\araa}{Ann.Rev.Astron.Astrophys.}

\newcommand{\bn}{{\bf n}}

\newcommand{\pr}{{\rm patchy reionization}}
\newcommand{\ds}{{\rm differential screening}}
\newcommand{\atau}{{\rm anisotropic optical depth}}
\newcommand{\pp}{{\rm paper}}

\newcommand{\hqe}{{\rm hybrid quadratic estimator}}
\newcommand{\hqes}{{\rm hybrid quadratic estimators}}

\newcommand{\tj}[6]{ \begin{pmatrix}
   #1 & #2 & #3 \\
   #4 & #5 & #6 
  \end{pmatrix}}

  \def\hp{ {\sc HEALPix}}

\begin{document}
\title{Searching for patchy reionization from cosmic microwave background with hybrid quadratic estimators}

\author{Chang Feng}
\email{changf@illinois.edu}
\affiliation{Department of Physics, University of Illinois at Urbana-Champaign, 1110 West Green Street, Urbana, Illinois, 61801, USA}

\author{Gilbert Holder}
\affiliation{Department of Physics, University of Illinois at Urbana-Champaign, 1110 West Green Street, Urbana, Illinois, 61801, USA}
\affiliation{Department of Astronomy, University of Illinois at Urbana-Champaign, 1002 West Green Street, Urbana, Illinois, 61801, USA}
\affiliation{Canadian Institute for Advanced Research, Toronto, Ontario M5G 1M1, Canada}

\begin{abstract}
We propose a \hqe\ to measure cross correlations between gravitational lensing of the cosmic microwave background (CMB) and differential screening effects arising from fluctuations in the electron column density, such as could arise from patchy reionization. The \hqes\ are validated by simulated data sets with both Planck and CMB-Stage 4 (CMB-S4) instrumental properties and found to be able to recover the cross-power spectra with almost no biases. We apply this technique to Planck 2015 temperature data and obtain cross-power spectra between gravitational lensing and differential screening effects. Planck data alone cannot detect the patchy-reionization-induced cross-power spectrum but future experiment like CMB-S4 will be able to robustly measure the expected signal and deliver new insights on reionization.
\end{abstract}

\maketitle

\section{Introduction}
Recombination of hydrogen atoms 380,000 years after the Big Bang left the early Universe with neutral hydrogen gas that was almost uniform but had small density fluctuations~\cite{2001ARA&A..39...19L}. These fluctuations seeded the population of the first stars that emitted ultraviolet (UV) radiation, by which electrons were stripped from neutral hydrogen atoms and scattered with cosmic microwave background (CMB) photons. Large inhomogeneities in the ionization fraction of the gas were created during the so-called epoch of reionization (EoR), leading to substantial variations in the CMB scattering optical depth. At later times there are additional modulations in the scattering optical depth that arise from fluctuations in the baryon density. These variations in the scattering optical depth cause secondary fluctuations in the CMB.
 
The secondary CMB anisotropies generated at the EoR are extremely weak but can create excess power in CMB temperature and polarization power spectra~\cite{2009PhRvD..79j7302D}. However, it is very difficult to detect the small amount of excess power from these secondary anisotropies in the presence of the substantial fluctuation power coming from the (Gaussian) primary fluctuations and instrument noise. To get sufficient sensitivities, higher order estimators for patchy reionization have been developed with three-point~\cite{2018arXiv180105396F} and four-point~\cite{cora1,2011arXiv1106.4313S} correlation functions. However, they either rely on a high-redshift large scale structure tracer which is hard to obtain or contain significant higher order biases that introduce extra uncertainties for the measurements. Although the auto-power spectrum of patchy reionization can be recovered after subtracting model-dependent biases, a Gaussian noise that is almost six orders of magnitude higher than the signal makes it hard to detect. 

Building on previous work~\cite{2018arXiv180105396F} which investigated the utility of cross-correlating a relatively noisy optical depth reconstruction with higher signal-to-noise tracers of large scale structure, we consider the CMB gravitational lensing as a high-redshift tracer and construct a cross correlation between gravitational lensing ($\phi$) and \ds\ ($\tau$) effects. This is essentially a four-point correlation function, using two-point estimators for $\phi$ and $\tau$, respectively. In this \pp, we study theoretical predictions of this cross correlation, and construct a hybrid quadratic estimator to extract such a signal from CMB data.

This paper is structured as follows: in Sec. II, we describe details of constructing the ``\hqe'' as well as various debiasing steps; in Sec. III, we validate the \hqe\ with simulations at noise levels appropriate for Planck and a CMB-S4-like experiment; we then apply this estimator to Planck 2015 temperature data in Sec. IV and conclude in Sec. V.\\

\section{Hybrid quadratic estimators}\label{sec_hqe}
The CMB with both gravitational lensing ($\phi$) and \ds\ ($\tau$) effects can be projected onto a unit sphere as
\begin{equation}
X(\bn)=\tilde X(\bn+\nabla\phi(\bn))e^{-\tau(\bn)},
\end{equation}
where $\bf n$ is a direction in the sky and $\tilde X$ is the unlensed CMB. The symbol $X$ refers to the lensed CMB (or the measured CMB) and it can be CMB temperature $T$ or polarization $Q\pm iU$. For CMB measurements, it is customary to express polarization measurements in terms of electric-like ($E$) and curl-like ($B$) modes. These are related to the Stokes parameters via $Q\pm iU=-\sum_{\ell m}(E_{\ell m}\pm iB_{\ell m}){}_{\pm2}Y_{\ell m}$ where ${}_{s}Y_{\ell m}$ are spin-$s$ spherical harmonics.

Noisy reconstructions of $\phi$ and $\tau$ ~\cite{2003PhRvD..67h3002O, cora1} are proportional to off-diagonal elements of the covariance matrix for two CMB modes $X_{\ell m}$ and $Z_{\ell'm'}$. In essence, a general form for any noisy reconstruction can be expressed as
\begin{eqnarray}
\Psi^{(c)}_{LM}(X_{\ell m},Z_{\ell'm'})&=&A^{(c)}_L\displaystyle\sum_{\ell m\ell'm'}(-1)^M\tj{\ell}{\ell'}{L}{m}{m'}{-M}\nonumber\\
&\times&g^{(c)}_{\ell\ell'}(L)X_{\ell m}Z_{\ell'm'},\label{qe}
\end{eqnarray}
where the modes $X_{\ell m}$ and $Z_{\ell'm'}$ can be $T_{\ell m}$ for temperature, and $E_{\ell m}$/$B_{\ell m}$ modes for polarization. The big bracket $(...)$ is the 3-$j$ Wigner symbol. Normalization functions $A_L$ and weighting functions $g_{\ell\ell'}(L)$ are given in ~\cite{2003PhRvD..67h3002O, cora1} for $\phi$ and $\tau$, respectively. The superscript ``(c)'' = $\{\phi,\tau\}$ and quadratic estimators for lensing and \atau\ are $\Psi^{(\phi)}=\hat\phi$ and $\Psi^{(\tau)}=\hat\tau$, respectively. A \hqe\ is defined as the cross power spectrum between these two reconstructed fields, $
{\hat C}_{\ell}^{\phi\tau}\delta_{\ell\ell'}\delta_{mm'}=\langle\Psi^{(\phi)}_{\ell m}\Psi^{(\tau,\ast)}_{\ell'm'}\rangle$. The weighting function $g^{(c)}_{\ell\ell'}(L)$ has a separable form and the \hqes\ can be expressed as products of filtered maps, thereby computation cost is significantly reduced~\cite{2016A&A...594A..15P}. 

We make a set of simulations $\{\hat\phi^{(\rm un)}, \hat\phi^{(\phi)},\hat \phi^{(\tau)}, \hat\phi^{(f)}, \hat\phi^{(\rm G)}\}$ for lensing potential and an analogous set of simulations for $\tau$. Here the superscripts ``(un)'', ``$(\phi)$'', ``$(\tau)$'', ``(f)'' and ``(G)'', refer to unlensed maps, lensed maps, maps only perturbed by $\tau$, maps with both $\phi$ and $\tau$ perturbations, and Gaussian simulations. A four-point correlation function is thus constructed from four CMB maps $X$, $Z$, $W$ and $V$:
\begin{eqnarray}
{\hat C}^{\phi\tau}_{\ell}&=&\langle[\hat\phi(X^{(s)},Z^{(s)})-\hat\phi(X^{(\tau)},Z^{(\tau)})]\nonumber\\
&\times&[\hat\tau(W^{(s)},V^{(s)})-\hat\tau(W^{(\phi)},V^{(\phi)})]\rangle,
\end{eqnarray}
where the superscript ``(s)'' refers to data or simulations. Mean field maps derived from simulations are subtracted from measured $\phi$ and $\tau$ reconstructions. An estimator-based bias [Eq. (\ref{tbias}) or (\ref{xbias})] is subtracted from a raw signal-plus-noise trispectrum in addition to the Gaussian and higher order biases ($N_1$). To account for any potential covariance mismatch between data and simulations, a realization dependent bias is subtracted from the measured raw trispectra instead of the simulated Gaussian bias [Eq. (\ref{RD})]~\cite{2014A&A...571A..17P,2015ApJ...810...50S}. We will verify in the next section that the higher order biases are negligible for the cross-power spectrum $\langle\phi\tau\rangle$ with different amplitudes predicted by a broad range of EoR models, but may not be negligible for the auto-power spectrum $\langle\tau\tau\rangle$ if its amplitude is too low.

\begin{figure}
\includegraphics[width=8cm, height=6cm]{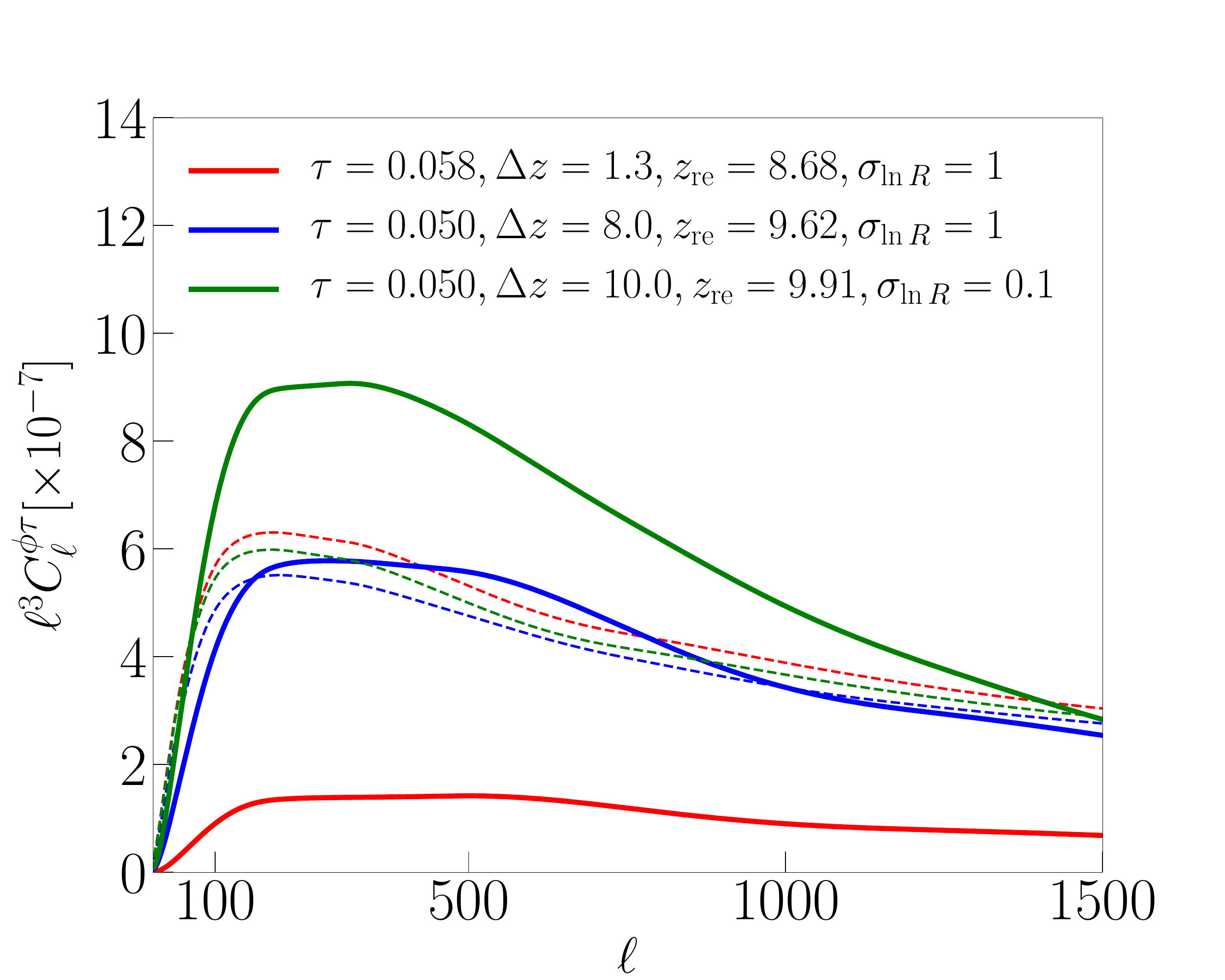}
\caption{Theoretical cross-correlations $\langle\phi\tau\rangle$ between CMB gravitational lensing and \ds\ effects. Contributions by patchy reionization (solid) and large scale structure (dashed) are shown with three EoR models. It is seen that the cross-power spectrum $\langle\phi\tau\rangle$ due to patchy reionization has a consistent shape but could have different amplitudes. The theoretical predictions in the green curves are calculated with assumptions that bubble size ($\bar R$) and bubble bias ($b_{\rm bubble}$) are free parameters as done in~\cite{cora1}. The parameter $z_{\rm re}$ in the legend is a characteristic redshift for an ionization history and is determined by the optical depth today and the duration of EoR. }\label{eortheory}
\end{figure}

\begin{figure*}
\includegraphics[width=18cm, height=7cm]{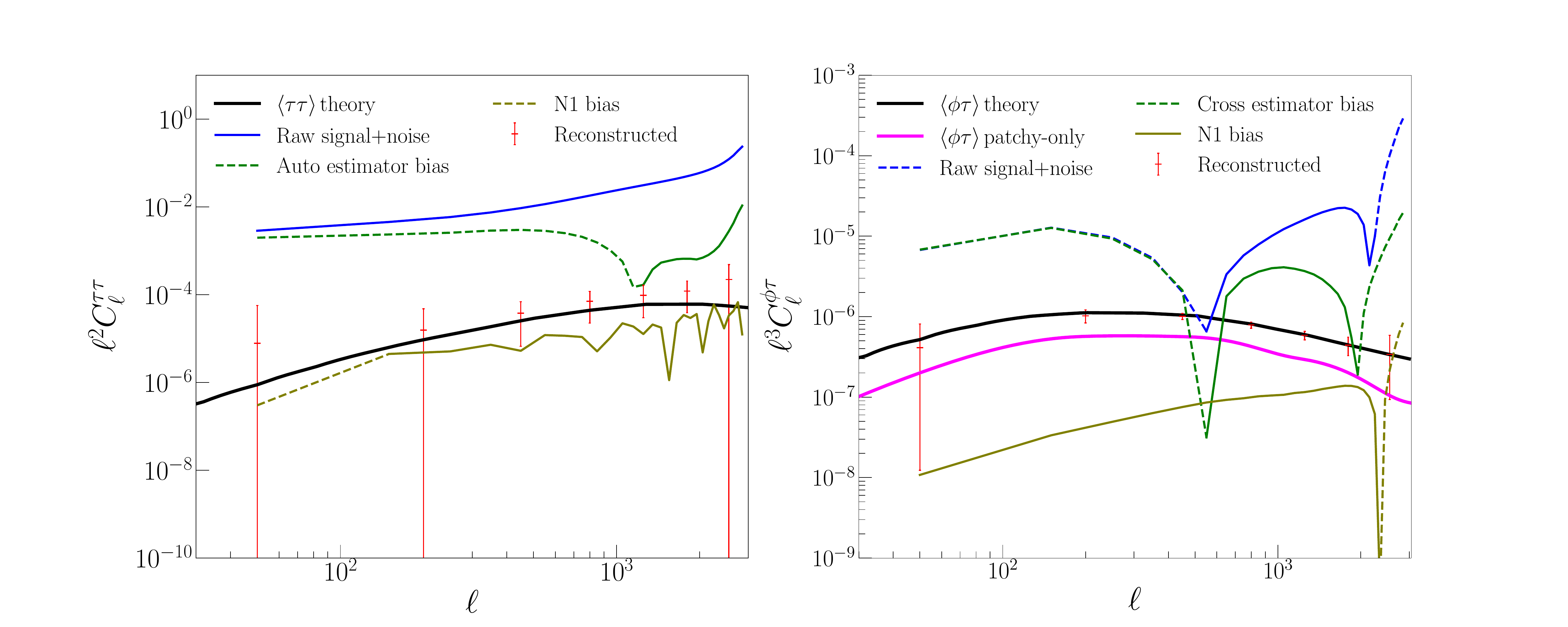}
\caption{Comparison of the auto-power spectrum estimator $\langle\tau\tau\rangle$ (left) and the cross-power spectrum estimator $\langle\phi\tau\rangle$ (right), i.e., the \hqe, with a CMB-S4-like data set. Here we only show detailed components for $\langle\tau(EB)\tau(EB)\rangle$ (left) and $\langle\phi(EB)\tau(EB)\rangle$ (right). The dashed portion is negative. Signal-to-noise ratios for the auto- and the cross-power spectra with the fiducial EoR model are $\sim 2\sigma$ and $\sim 22\sigma$, respectively. }\label{comparison}
\end{figure*}

\begin{figure*}
\includegraphics[width=18cm, height=6cm]{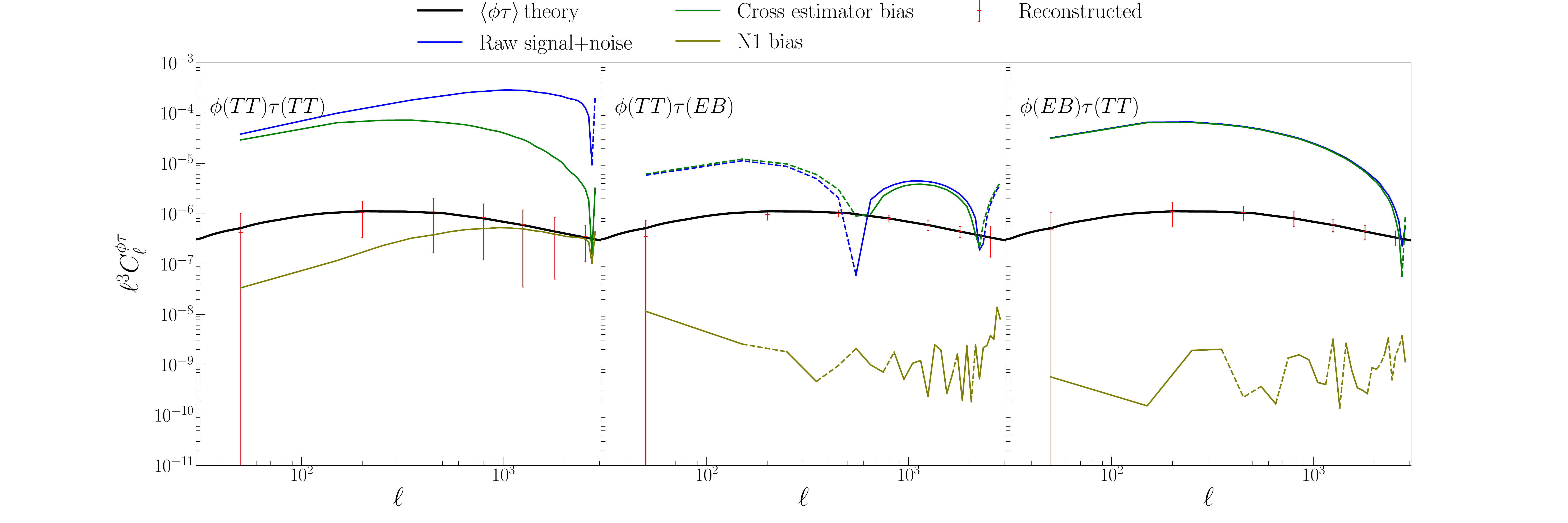}
\caption{Validations of the \hqes\ with a CMB-S4-like data set. Simulations are made at \hp\ resolution $N_{\rm{side}}=2048$. The signal-to-noise ratios are $4\sigma$ for $\langle\phi(TT)\tau(TT)\rangle$, $14\sigma$ for $\langle\phi(TT)\tau(EB)\rangle$, $9\sigma$ for $\langle\phi(EB)\tau(TT)\rangle$ and $22\sigma$ for $\langle\phi(EB)\tau(EB)\rangle$ [Fig. \ref{comparison} (right)], respectively. The CMB modes within $200<\ell<3000$ are used for the reconstructions. The dashed portion is negative.}\label{crossvalidcmbs4}
\end{figure*}

\section{Numerical validations} \label{sec_numeric}
We simulate CMB temperature and polarization for a next generation experiment like CMB Stage-4 (CMB-S4) with noise levels at $\Delta_T=1\mu\rm{K}\mbox{-}\rm{arcmin}$ and $\Delta_P=\sqrt{2}\mu\rm{K}\mbox{-}\rm{arcmin}$. We assume a Gaussian beam with $1'$ full width at half maximum (FWHM) $\theta$ and a full-sky coverage. The unlensed CMB power spectra are calculated by CAMB~\footnotemark[1]\footnotetext[1]{https://camb.info}{}. The lensed CMB simulations are made by \textit{Taylens}~\cite{2013JCAP...09..001N} at \hp~\cite{hp} resolution $N_{\rm side}=2048$. Patchy-reionization-related theory power spectra $C^{\tau\tau}_{\ell}$ and $C^{\phi\tau}_{\ell}$ are calculated using the halo model formalism ~\cite{2002PhR...372....1C, 2013ApJ...779..124M}, in which the CAMB's $\tanh$ reionization model is used~\cite{2008PhRvD..78b3002L} and the size of ionizing bubbles is assumed to satisfy a logarithmic distribution~\cite{2009ApJ...690..252L}.  The bubble size and bubble bias are self-consistently solved for a given reionization history. The fiducial parameters for the EoR are $\{\tau=0.05,\Delta z=8, \sigma_{\ln R}=1\}$, where $\tau$ is the optical depth today, $\Delta z$ is duration of the EoR and $\sigma_{\ln R}$ is the variance of the ionizing bubbles. A representative plot for the cross-power spectrum $C^{\phi\tau}_{\ell}$ is shown in Fig. \ref{eortheory}, where a calculation made by a toy model, assuming the bubble size and bubble bias are free parameters, is also shown in green for comparison~\cite{cora1}. From a family of cross-power spectra, it is seen that the large-scale-structure (LSS) generated piece has a weak dependence on the EoR models, but the one generated by \pr\ has a very uncertain amplitude.

We made simulations for $\phi_{\rm input}$ and $\tau_{\rm input}$ by performing the Cholesky decomposition of the theoretical covariance between them. These simulations are used as input for generation of mock CMB data. For a CMB-S4-like experiment, we make white noise simulations and convolve mock CMB maps with the beam profile. We apply the standard quadratic estimators $\phi$ and $\tau$ to the mock CMB data and create reconstructed $\hat\phi$ and $\hat\tau$ maps. We validate that the cross-power spectra $\langle\hat\phi\phi_{\rm input}\rangle$ and $\langle\hat\tau\tau_{\rm input}\rangle$ are consistent with input power spectra $C_{\ell}^{\phi\phi}$ and $C_{\ell}^{\tau\tau}$.

\begin{figure}
\includegraphics[width=9cm, height=7cm]{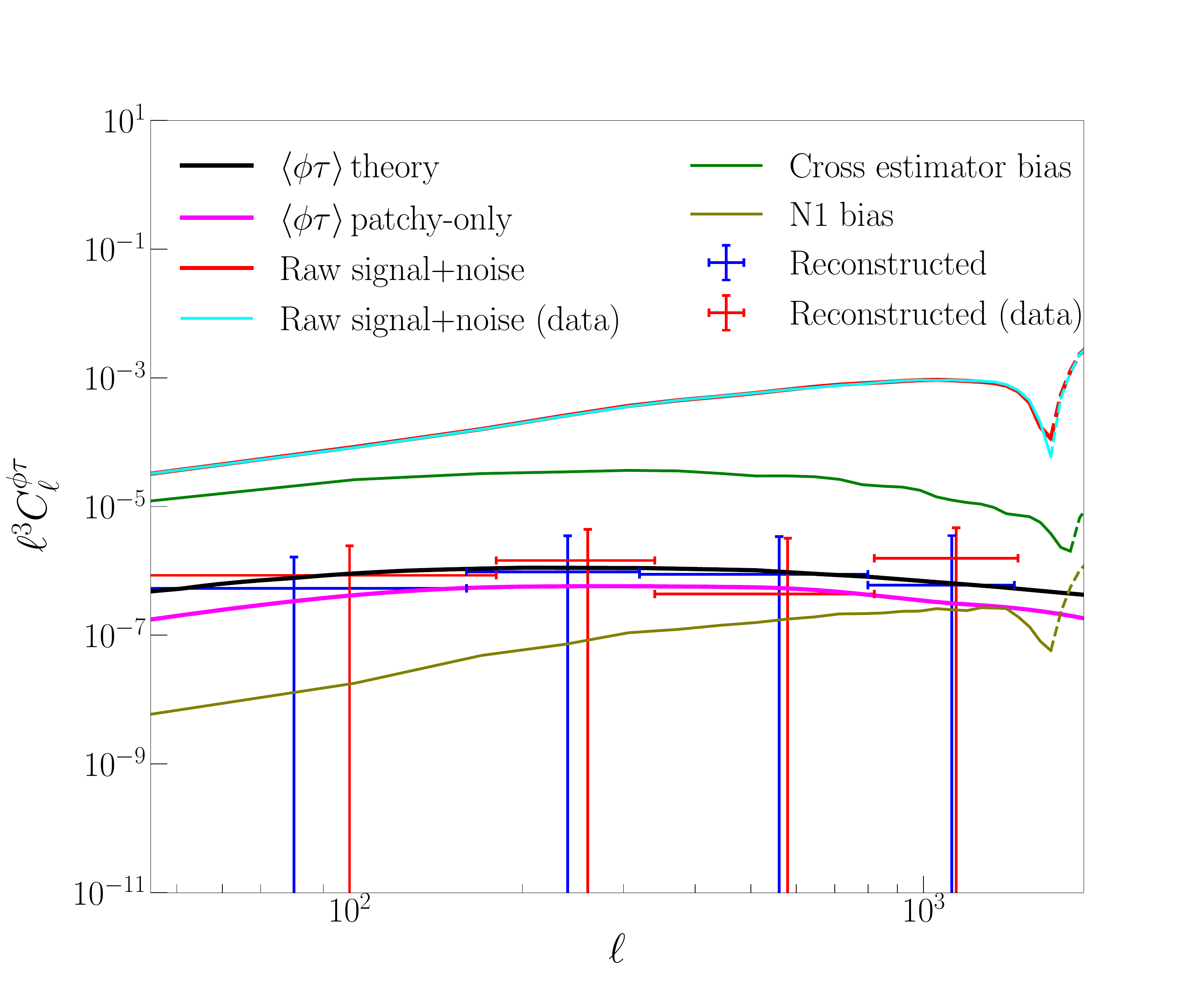}
\caption{Cross-power spectrum $\langle\phi\tau\rangle$ for Planck simulations and data. Planck full focal plane simulations FFP8.1 are used for the noise component. The dashed portion is negative.}\label{crossdataRAW}
\end{figure}

\begin{figure*}
\includegraphics[width=8cm, height=6cm]{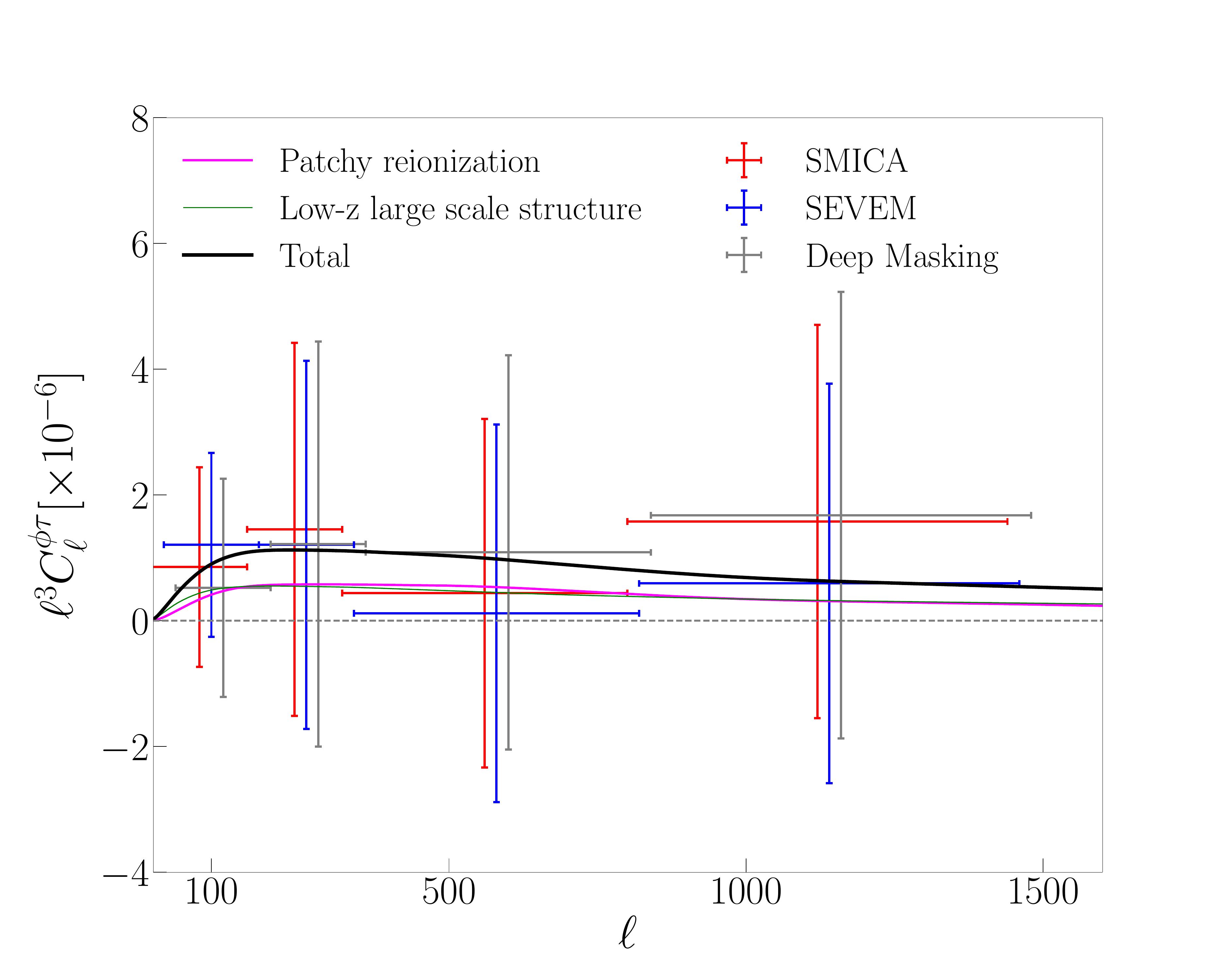}
\includegraphics[width=8cm, height=6cm]{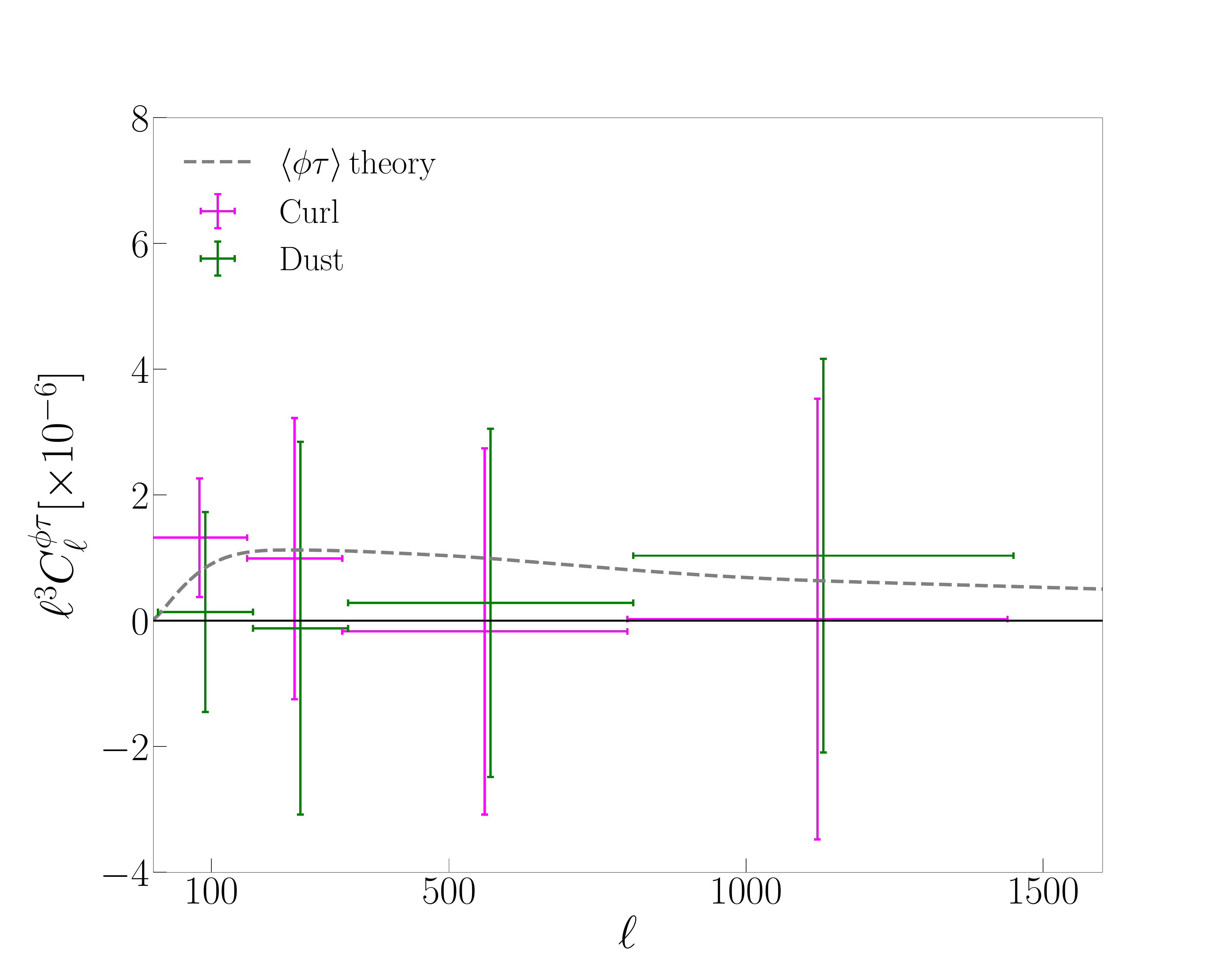}
\caption{ Left: measured cross-power spectrum $\langle\phi\tau\rangle$ from 2015 Planck component-separated temperature maps SMICA and SEVEM. Right: systematic tests for the Planck SMICA data. }\label{crossdata}
\end{figure*}

After subtracting the biases calculated using simulations, the cross-power spectrum $\langle\phi\tau\rangle$ can be recovered as Fig. \ref{comparison} shows. However, the auto-power spectrum $\langle\tau\tau\rangle$ can be recovered at a level that the residual bias is smaller than the statistical uncertainty. Also, the residual bias of the auto-power spectrum is model dependent and would become negligible if the amplitude of the input model $\langle\tau\tau\rangle$ is lower than our nominal model. The sensitivity of the reconstructed auto-power spectrum is much lower than the cross-power spectrum $\langle\phi\tau\rangle$ that is derived from the same reionization model. Also, the auto-power spectrum $\langle\tau\tau\rangle$ may strongly depend on the higher order bias ($N_1$) that must be determined from the desired patchy reionization signal, and model uncertainty will complicate interpretations of data measurements. On the other hand, the numerical simulations show that the reconstructed cross-power spectra $\langle\phi\tau\rangle$ have negligible higher order biases and a precise knowledge of the reionization model is not required. We validated this by repeating the same calculation with a minimum EoR model [red curves in Fig. (\ref{eortheory})] and we found that the auto- and cross-power spectra are both correctly recovered, and the higher order biases are negligible for the cross but not for the auto. This can be understood intuitively: the cross-power spectrum $\langle\phi\tau\rangle$ only relies on ${\mathcal O}(\tau)$ not ${\mathcal O}(\tau^2)$, so it is less dependent of the poorly known reionization model. 

In Fig. \ref{crossvalidcmbs4}, we show various components generated by the $\langle\phi\tau\rangle$ \hqes\ with CMB temperature and polarization simulations and validate that the \hqes\ can recover the cross-power spectrum $C^{\phi\tau}_{\ell}$ from different CMB combinations. The calculation shows that, for a CMB-S4-like experiment, most of the signal-to-noise comes from the polarization trispectrum $\langle EBEB\rangle$ and the total signal-to-noise ratio would reach a $\sim 30\sigma$ level if all the cross-power spectra are combined. This cross-spectrum is a four-point correlation function resembling the reconstructed lensing power spectrum, but applies different filters to the various mode pairs. It is still an internal measurement that only uses CMB data as an external large scale structure tracer is not required. 

Signal-to-noise ratios (SNRs) for the cross-power spectra $\langle\phi\tau\rangle$ are roughly one order of magnitude higher than the auto-power spectra $\langle\tau\tau\rangle$. The higher SNRs of the \hqe\ can be easily understood by introducing an equivalent reconstruction noise $N_0^{\phi\tau}\propto\sqrt{N_0^{\phi\phi}N^{\tau\tau}_0}$, from which the SNR can be approximated as a scaling law $\sqrt{N^{\tau\tau}_0/N^{\phi\phi}_0}$. Here $N_0$ refers to Gaussian biases of $\phi$ and $\tau$.\\

\section{Measurements from Planck 2015 temperature data} \label{sec_planck}

\begin{figure}
\includegraphics[width=8cm, height=6.4cm]{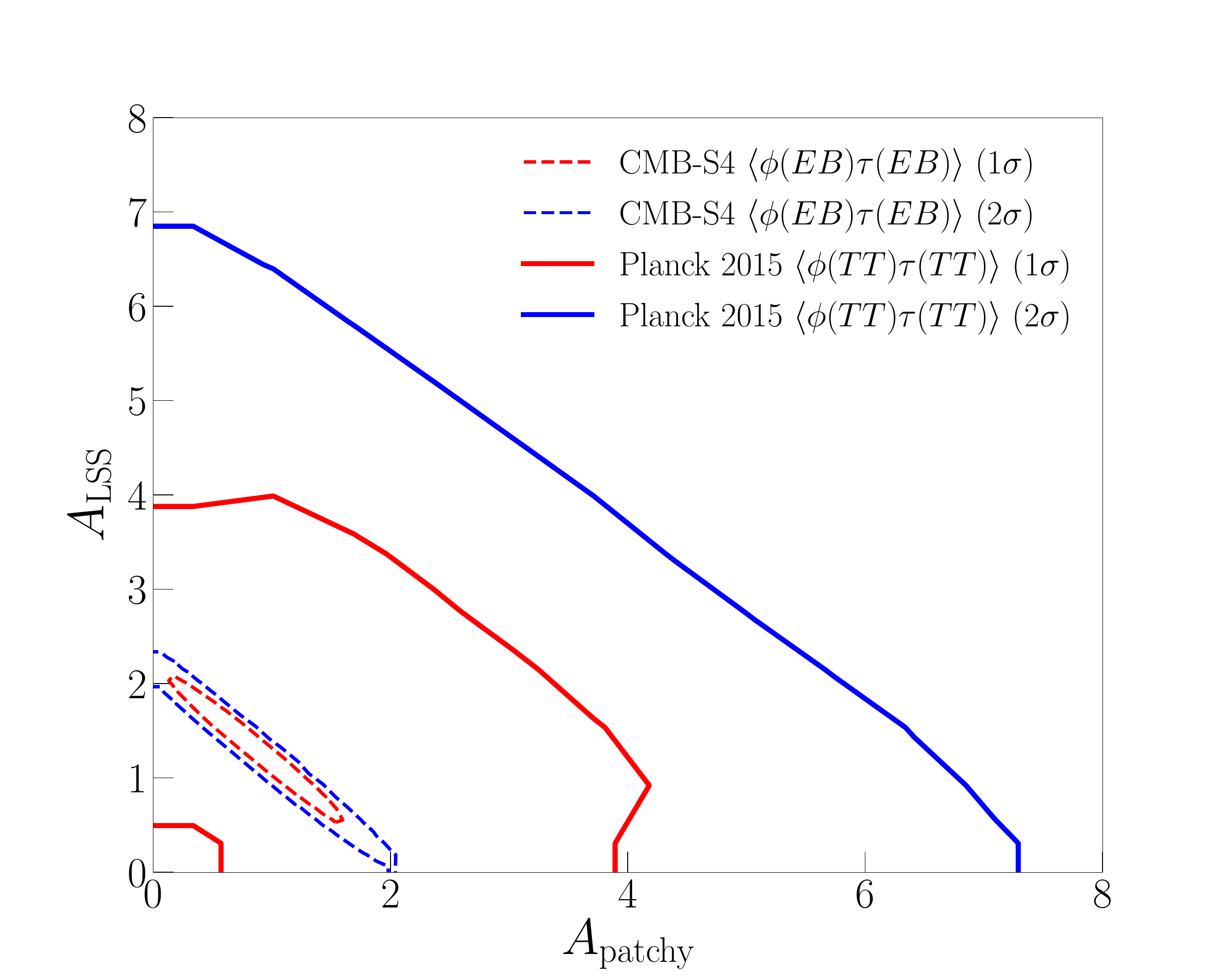}
\caption{Constraints on component amplitudes for the cross-power spectra $\langle\phi\tau\rangle$ generated by patchy reionization and large scale structure, respectively. Planck and CMB-S4-like constraints are shown in solid and dashed lines, respectively.}\label{eor}
\end{figure}

To measure a cross-power spectrum $\langle\phi\tau\rangle$ from data with the \hqe, we use the 2015 Planck SMICA temperature data~\cite{2016A&A...594A...9P}. It is foreground-cleaned and is less contaminated by foregrounds than single frequency maps. The beam profile of SMICA data can be well approximated as a $5'$ Gaussian beam. An analysis mask is applied to remove point sources and Galactic plane, resulting in a sky fraction $f_{\rm sky}=0.67$. The mask is apodized with $60'$ Gaussian beam to suppress mode-mixing induced by the sky cut. To reduce potential systematic effects and dust contamination, we filter out the data below $\ell_{\rm min}=200$. We also set $\ell_{\rm max}=2048$ to avoid potential high-$\ell$ foreground residuals. We use the Planck full focal plane (FFP) simulation, i.e., FFP8.1 SMICA noise simulations~\cite{2016A&A...594A..12P}~\footnote[2]{https://crd.lbl.gov/departments/computational-science/c3/c3-research/cosmic-microwave-background/cmb-data-at-nersc/}{}, and have checked that the noise power spectra from SMICA component-separation simulations can precisely match the measured ones.

The reconstructions with the \hqes\ are performed with a full suite of Planck simulations. We have repeated the same validation procedures, as for the CMB-S4-like experiment in the previous section, for the Planck 2015 data. We show different components of the cross-power spectrum $\langle\phi\tau\rangle$ for the Planck 2015 simulations in Fig. \ref{crossdataRAW}, where a bias-free reconstruction (blue band powers) of the cross-power spectrum $\langle\phi\tau\rangle$ is achieved after subtracting simulation-derived biases. We only use the Planck temperature data, as the polarization data are too noisy to be used for this analysis. 

In Fig. \ref{crossdata} (left), we show the measured cross-power spectrum $\langle\phi\tau\rangle$. We compare the theoretical models to the measurement and calculate a $\chi^2$ from
\begin{equation}
\chi^2=\displaystyle\sum_{bb'}(\hat C_{b}-C^{\rm th}_{b})\textbf{C}^{-1}_{bb'}(\hat C_{b'}-C^{\rm th}_{b'}).
\end{equation}
Here $\textbf{C}$ is the covariance matrix between different band powers and $\hat C_{b}$ is the measured $\langle\phi\tau\rangle$ band power at band $b$. The theoretical band power is decomposed into two parts: $C^{\rm th}_{b}=\mathcal{A}_{\rm patchy}C^{\rm patchy}_{b}+\mathcal{A}_{\rm LSS}C^{\rm LSS}_{b}$, where $C^{\rm patchy}_{b}$ is the theoretical cross-power spectrum created by \pr\ (blue solid line in Fig. \ref{eortheory}), and $C^{\rm LSS}_{b}$ is the cross-power spectrum due to the low-$z$ large scale structure (LSS, blue dashed line in Fig. \ref{eortheory}). This simple parametrization is chosen to avoid degeneracy of the EoR parameters which are still difficult to be constrained from this measurement.

The anti-correlation between the two amplitudes is shown in Fig. \ref{eor} if both amplitudes are varied. Given the fact that the LSS contribution $\mathcal{A}_{\rm LSS}$ does not vary too much as seen from the theoretical calculations in Fig. \ref{eortheory}, we fix $\mathcal{A}_{\rm LSS}$ to be 1 and find that $\mathcal{A}_{\rm patchy}<6$ at 95\% confidence level. Compared to the fiducial model ($\mathcal{A}_{\rm patchy}$=1, $\mathcal{A}_{\rm LSS}=1$), the SMICA measurement (red data points) has an overall amplitude $\mathcal{A}<3$ at 95\% confidence level, and the $\chi^2/\rm{dof}$ is found to be 0.22/4. We replace the SMICA map by SEVEM and use its FFP8 noise simulations. The same analysis is performed and the measurement (in blue) is consistent with the one from SMICA data. Moreover, we test a deeper masking scheme for the SMICA data and find that a more expansive sky cut with $f_{\rm sky}\sim$ 0.57 results in a negligible change to the cross-power spectrum [Fig. \ref{crossdata} (left)]. 

We construct a curl null estimator~\cite{2005PhRvD..71l3527C} for the lensing potential and use it to measure the cross-power spectrum $\langle\phi^{\rm curl}\tau\rangle$ with an expected signal of zero. In Fig. \ref{crossdata} (right), it is found to be consistent with zero and the PTE is 79\%. Given the fact that the SMICA data is foreground-cleaned and only the temperature modes within $200<\ell<2048$ are used, contaminating power of dust and foregrounds on the cross-power spectrum $\langle\phi\tau\rangle$ should be negligible. To verify this, we investigate the maximum impact with no dust isolation and propagate a Planck thermal dust map ``$\rm{COM}\_\rm{CompMap}\_\rm{dust}\mbox{-}\rm{commrul}\_2048\_\rm{R1.00.fits}$'' through the same analysis pipeline. The dust-induced trispectrum $\langle\phi\tau\rangle$ shown in green in Fig. \ref{crossdata} (right) is consistent with zero. 

From the Planck temperature data, the cross-power spectrum $\langle\phi\tau\rangle$ is not detected (at $\sim1\sigma$ statistical significance level), but will be detected by the CMB-S4-like experiment, and the parameter space on the $\mathcal{A}_{\rm patchy}\mbox{--}\mathcal{A}_{\rm LSS}$ plane will be significantly reduced as seen from Fig. \ref{eor}, where only simulated band powers from $\langle\phi(EB)\tau(EB)\rangle$ are taken into account.

 \section{Conclusions} \label{con}
In this \pp, we propose a \hqe\ formalism for the study of fluctuations in the \atau, such as would be induced by \pr. We make mock data sets for Planck and CMB-S4-like experiments and numerically validate that the cross-power spectrum $\langle\phi\tau\rangle$ can be correctly recovered from the CMB data alone. Moreover, we measure a $\langle\phi\tau\rangle$ cross-power spectrum from Planck 2015 temperature data and obtain a new upper bound for \pr. Various systematic and foreground tests are performed and their effects are found to be negligible. For the next-generation CMB experiments, both temperature and polarization data can be used to measure the cross-power spectra $\langle\phi\tau\rangle$ with much higher signal-to-noise ratios, and the signal of patchy reionization in future experiments will be detected from CMB alone, extending CMB science to a new regime when the first stars and galaxies formed.

\section{Acknowledgments}
This research was supported in part by Perimeter Institute for Theoretical Physics. Research at Perimeter Institute is supported by the Government of Canada through the Department of Innovation, Science, and Economic Development, and by the Province of Ontario through the Ministry of Research and Innovation. This research is supported by the Brand and Monica Fortner Chair, and is part of the Blue Waters sustained-petascale computing project, which is supported by the National Science Foundation (awards OCI-0725070 and ACI-1238993) and the state of Illinois. Blue Waters is a joint effort of the University of Illinois at Urbana-Champaign and its National Center for Supercomputing Applications. We also acknowledge the use of the \hp~\cite{hp} package.

\bibliography{hybrid}

\clearpage
\newpage

\onecolumngrid

\begin{appendix}

\begin{center}
{\bf SUPPLEMENTAL MATERIAL}
\end{center}
In this supplementary material, we describe various trispectrum components for the patchy-reionization related estimators, including realization dependent biases, estimator-based biases and higher order ($N_1$) biases. 

\section{Trispectrum components for patchy-reionization related estimators}
The weighting functions for estimators of gravitational lensing and anisotropic optical depth are given in Table \ref{wf}, where the $F$ and $J$ matrix are multipole-weighted 3-$j$ Wigner symbols~\cite{2003PhRvD..67h3002O, cora1}. Symbols $\tilde C_{\ell}$ and $C_{\ell}$ refer to unlensed and noise-included CMB power spectra.

A four-point correlation function is constructed from four CMB maps $X$, $Z$, $W$ and $V$. $\{X,Z,W,V\}=\{T,E,B\}$. We make simulations for a map set for lensing potential $(\phi^{(\rm un)}, \phi^{(\phi)}, \phi^{(\tau)}, \phi^{(\rm f)}, \phi^{(\rm G)})$. Here the superscripts ``(un)'', ``$(\phi)$'', ``$(\tau)$'', ``(f)'' and ``(G)'', refer to unlensed, $\phi$-only, $\tau$-only, $\phi$+$\tau$, and Gaussian simulations. The same map set is also made for $\tau$ reconstruction. In the following, operators $\{A, B\}=\{\phi,\tau\}$. Generally, the four-point correlation function is expressed as 
\begin{equation}
C_{\ell}^{A\times B,cd}=\langle A(X^{(c)},Z^{(c)})B(W^{(d)},V^{(d)})\rangle,
\end{equation}
where the superscript ``$c$'' or ``$d$'' refers to a specific type of CMB map in the map set. 

The realization dependent (RD) bias is calculated from
\begin{eqnarray}
N_{0,\ell}^{A\times B,cd}(XZ;WV)&=&\langle[A(X^{(D)},Z^{(c)})+A(X^{(c)},Z^{(D)})][B(W^{(D)},V^{(c)})+B(W^{(c)},V^{(D)})]\rangle\nonumber\\
&-&\langle A(X^{(c)},Z^{(c')})[B(W^{(d)},V^{(d')})+B(W^{(d')},V^{(d)})]\rangle.\label{RD}\nonumber\\
\end{eqnarray}
Here the superscript ``($D$)'' refers to the data. We make two map realizations labeled as $(X,Z,W,V)$ and $(X',Z',W',V')$.

The estimator for the anisotropic optical depth $\langle\tau\tau\rangle$ is defined as
\begin{eqnarray}
{\hat C}^{\tau\tau}_{\ell}
&=&\langle[\tau(X^{(f)},Z^{(f)})-\tau(X^{(\phi)},Z^{(\phi)})][\tau(W^{(f)},V^{(f)})-\tau(W^{(\phi)},V^{(\phi)})]\rangle.\label{tauest}
\end{eqnarray}
In the following, the average symbol $\langle\rangle$ and subscript $\ell$ are omitted for brevity. Eq. \ref{tauest} can be decomposed into a few components. The raw signal is
\begin{equation}
\tilde C^{\tau\tau}=\tau(X^{(f)}, Z^{(f)})\tau(W^{(f)}, V^{(f)}),
\end{equation}
and the raw lensing-induced trispectrum is
\begin{eqnarray}
\tilde C^{\tau\tau, \phi}&=&\tau(X^{(f)},Z^{(f)})\tau(W^{(\phi)},V^{(\phi)})+\tau(X^{(\phi)},Z^{(\phi)})\tau(W^{(f)},V^{(f)})-\tau(X^{(\phi)},Z^{(\phi)})\tau(W^{(\phi)},V^{(\phi)}),\label{tbias}
\end{eqnarray}
which is related to a pure lensing bias as $C^{\tau\tau,\phi}=\tilde C^{\tau\tau, \phi}-C^{\tau\tau, G}$, and $C^{\tau\tau, G}$ is the Gaussian bias.

The $N_1$ bias for $\tau\tau$ ($A=\hat\tau$, $B=\hat\tau$, $c$ = $d$ =``($\tau$)'') is
\begin{eqnarray}
N_{1,\ell}^{A\times B,cd}(XZ;WV)&=&\langle A(X^{(c)},Z^{(c')})[B(W^{(d)},V^{(d')})+B(W^{(d')},V^{(d)})]\rangle\nonumber\\
&-&\langle A(X^{(c)},Z^{(c'')})[B(W^{(d)},V^{(d'')})+B(W^{(d'')},V^{(d)})]\rangle,\nonumber\
\end{eqnarray}
where two realizations $X,Z,W,V$ and $X',Z',W',V'$ are created from the same $\phi$ or $\tau$ but different CMB and noise realizations.  

With the components defined above, the reconstructed signal is expressed as 
\begin{eqnarray}
{\hat C}^{\tau\tau}_{\ell}
&=&\tilde C_{\ell}^{\tau\tau}-C_{\ell}^{\tau\tau,\phi}-N_{0,\ell}-N_{1,\ell},
\end{eqnarray}
where the RD is subtracted instead of the Gaussian bias $N_0$ from simulations for the data.

The estimator of the cross-power spectrum $\langle\phi\tau\rangle$ is defined as
\begin{eqnarray}
{\hat C}^{\phi\tau}_{\ell}&=&\langle[\phi(X^{(f)},Z^{(f)})-\phi(X^{(\tau)},Z^{(\tau)})][\tau(W^{(f)},V^{(f)})-\tau(W^{(\phi)},V^{(\phi)})]\rangle.
\end{eqnarray}

The raw trispectrum is 
\begin{equation}
{\tilde C}^{\phi\tau}=\phi(X^{(f)},Z^{(f)})\tau(W^{(f)},V^{(f)}),
\end{equation}
and the lensing-induced raw trispectrum is
\begin{eqnarray}
\tilde  C^{\phi\tau,\phi}&=&\phi(X^{(\tau)},Z^{(\tau)})\tau(W^{(f)},V^{(f)})+\phi(X^{(f)},Z^{(f)})\tau(W^{(\phi)},V^{(\phi)})-\phi(X^{(\tau)},Z^{(\tau)})\tau(W^{(\phi)},V^{(\phi)}),\nonumber\\\label{xbias}
\end{eqnarray}

\begin{table*}
\caption{The weighting functions for lensing ($\phi$) and patchy reionization ($\tau$) quadratic estimators~\cite{2003PhRvD..67h3002O,cora1}}
  \begin{tabular}{ccccc}
    \hline
&$f^{\phi}_{\ell L\ell'}$&$f^{\tau}_{\ell L\ell'}$&$g^{\phi}_{\ell\ell'}(L)$&$g^{\tau}_{\ell\ell'}(L)$\\
  \hline
$TT$&${\tilde C}^{TT}_{\ell} {}_0F_{\ell'L\ell}+{\tilde C}^{TT}_{\ell'}{}_0F_{\ell L\ell'}$&$({\tilde C}^{TT}_{\ell}+{\tilde C}^{TT}_{\ell'})J^{\ell L\ell'}_{000}$&$\frac{f^{\phi}_{\ell L\ell'}}{2C^{TT}_{\ell}C^{TT}_{\ell'}}$&$\frac{f^{\tau}_{\ell L\ell'}}{2C^{TT}_{\ell}C^{TT}_{\ell'}}$\\
$EB$&${\tilde C}^{EE}_{\ell} {}_2F_{\ell'L\ell}$&${\tilde C}^{EE}_{\ell} (J^{\ell'L\ell}_{-202}-J^{\ell L\ell'}_{20-2})/(2i)$&$\frac{f^{\phi}_{\ell L\ell'}}{C^{EE}_{\ell}C^{BB}_{\ell'}}$&$\frac{f^{\tau}_{\ell L\ell'}}{C^{EE}_{\ell}C^{BB}_{\ell'}}$\\
    \hline
  \end{tabular}\label{wf}
\end{table*}
from which a pure lensing bias is defined as $C^{\phi\tau,\phi}=\tilde C^{\phi\tau, \phi}-C^{\phi\tau, G}$, and $C^{\phi\tau, G}$ is the Gaussian term estimated from Gaussian realizations.

The $N_1$ bias for the cross-power spectrum is
\begin{eqnarray}
N_{1,\ell}=N_{1,\ell}^{A\times B,(ff)}-N_{1,\ell}^{A\times B,(f\phi)}-[N_{1,\ell}^{A\times B,(\tau f)}-N_{1,\ell}^{A\times B,(\tau\phi)}],
\end{eqnarray}
where $A=\hat\phi$, $B=\hat\tau$, and each term has a general form
\begin{eqnarray}
N_{1,\ell}^{A\times B,cd}(XZ;WV)&=&\langle A(X^{(c)},Z^{(c')})[B(W^{(d)},V^{(d')})+B(W^{(d')},V^{(d)})]\rangle.\nonumber\\
\end{eqnarray}

With the components defined above, the reconstructed signal is expressed as 
\begin{eqnarray}
{\hat C}^{\phi\tau}_{\ell}=\tilde C_{\ell}^{\phi\tau}-C_{\ell}^{\phi\tau,\phi}-N_{0,\ell}-N_{1,\ell},
\end{eqnarray}
where the RD is subtracted instead of the Gaussian bias $N_0$ from simulations for the data.

\end{appendix}

\newpage
\end{document}